\documentclass[twocolumn,showpacs,showkeys,preprintnumbers,amsmath,amssymb]{revtex4}
\usepackage{dcolumn}
\usepackage{bm}
\usepackage[dvips]{graphicx}
\begin{document}
\title{Proximity and Josephson effects in superconductor \\
- antiferromagnetic Nb / $\gamma$-Fe$_{50}$Mn$_{50}$ heterostructures}
\author{C. Bell, E. J. Tarte, G. Burnell, C. W. Leung, D.-J. Kang, M. G. Blamire}
\affiliation{Materials Science Department, IRC in Superconductivity and IRC in Nanotechnology, \\
University of Cambridge, United Kingdom}
\date{\today}
\begin{abstract}
We study the proximity effect in superconductor (S),
antiferromagnetic (AF) bilayers, and report the fabrication and
measurement of the first trilayer S/AF/S Josephson junctions. The
disordered f.c.c. alloy $\gamma$-Fe$_{50}$Mn$_{50}$ was used as the
AF, and the S is Nb. Micron and sub-micron scale junctions were
measured, and the scaling of $J_C (d_{AF})$ gives a coherence
length in the AF of 2.4 nm, which correlates with the coherence length due
to suppression of $T_C$ in the bilayer samples. The diffusion
constant for FeMn was found to be $1.7 \times 10^{-4}$
m$^2$s$^{-1}$, and the density of states at the Fermi level was also
obtained. An exchange biased FeMn/Co bilayer confirms the AF
nature of the FeMn in this thickness regime. 
\end{abstract}
\pacs{74.45.+c  75.50.Ee  85.25.Cp  85.70.Kh}
\keywords{Josephson junction, antiferromagnet, proximity effect,
exchange bias}
\maketitle
\section{\label{AFMSC}Introduction}
From the earlist work of Hauser et al \cite{hauser} there has been
considerable interest in the proximity effect between thin film superconductors (S) with both
ferromagnetic (F) and antiferromagnetic (AF) materials. More recently 
Nb/Cr multilayers \cite{Cheng}, Nb/CuMn (spin-glass) systems have been studied \cite{attanC}, as have
Cr/V/Cr trilayers \cite{hubener}. The recently renewed interest
in the S/F proximity effect due to the oscillating order parameter
in the F layer and the so-called $\pi$-shift, recently found
experimentally \cite{ryazanov1, kontos}, has not led to complimentary
experiments in S/AF heterostructures. As has been pointed out by Krivoruchko
\cite{Krivoruchko}, the nesting features of Fermi surfaces of the
so-called band-antiferromagnets destroys the symmetry in momentum
space, similar to the splitting of the Fermi surface in the F
case. Therefore a band AF heavily suppresses superconductivity, but without the oscillating
order parameter found in the F case which is necessary to realise
$\pi$-junctions.

There has been much theoretical literature concerning S/F
heterostructures (bilayer, trilayers, and multilayers: for a
review see \cite{izyREVIEW}). The various effects on the
superconducting critical temperature, field and current density,
($T_C$, $H_C$ and $J_C$), of the parallel and anti-parallel
configurations of the F layers is of great interest. Various
theoretical predictions have been made for $J_C$ enhancement in
the case of S/F/X/F/S Josephson junctions, when the F layers have
their moments switched from parallel to anti-parallel, with large
effects for X = insulator
(I),\cite{BergeretPRL,GolubovJETP,krivoruchkoSFIFS}, more weakly
in the case of X = normal (dirty) metal (N) \cite{ChtchelJETP}.
Spin torque on the F layers due to the Josephson current 
has been predicted for X = N \cite{WaintalPRB}. To achieve these
`spin-active' junctions, techniques can be borrowed from the
magnetics community in the fabrication of spin-valve devices. In
these cases the anti-parallel alignment is achieved either by the use
of two materials with different coercive fields, or by using an AF to
`pin' (exchange bias) one of two otherwise identical F layers.
In the latter case this can be done by field cooling through the blocking temperature
(which is $\le T_N$, the N\'eel temperature), which increases the
coercive field, and shifts the centre of the magnetic hysteresis loop to a
non-zero applied field. The latter technique was used in the F/S/F
trilayers studied in \cite{guPRL}. If this technique is to be
applied to the case of S/F/X/F/S Josephson junctions, it is
crucial to understand the Josephson effect through an AF, as the
device is built up layer by layer.

The effect of magnetic and non-magnetic impurities in the barriers
of S/N/S junctions has been previously studied - including the
spin glasses CuMn, and AgMn \cite{Yang1984, Niemeyer1979}, and
CuNi \cite{paterson1979}. In these cases the barrier
thicknesses were of the order of 100 nm or thicker, (since the
impurity concentrations were relatively small: for example a maximum of 4.6 at.\% Mn in the
case of CuMn in \cite{Niemeyer1979}). To our knowledge no
measurements have ever been made of AF Josephson junctions and,
in particular, with the AF $\gamma$-Fe$_{50}$Mn$_{50}$. In this
paper we present bilayer $T_C$ measurements of the proximity
effect between FeMn and Nb, and the first measurement of the
Josephson effect through an AF. These measurements enable the
coherence length in the AF to be found, and hence the diffusion
constant, and in addition, the density of states at the Fermi level of the FeMn to
be calculated.

\section{Experimental details}
All films were deposited on (100) oxidised silicon substrates by
d.c. magnetron sputtering at 0.5 Pa, in an in-plane magnetic field
$\mu_0 H \sim 40$ mT. The sputtering system was cooled with liquid
nitrogen and had a base pressure better than $3 \times
10^{-9}$ mbar. The system was fitted with a load lock, which minimizes run
to run variation by keeping the targets under constant vacuum.
Deposition rates were of the order of 0.08 nm/s for the Cu, Co and
FeMn targets, and 0.03 nm/s for Nb. Film thicknesses were controlled by
varying the exposure time under the magnetron targets.
For all samples containing FeMn, a 5 nm underlayer of Cu was grown,
in order to achieve the required f.c.c. AF $\gamma$-FeMn phase \cite{tsang}.

For the trilayer devices, a Nb/Cu/FeMn/Nb sandwich was grown
{\it in situ}; FeMn thickness, $d_{\mathrm{FeMn}}$ was
in the range $2 - 6$ nm, both Nb thicknesses were 150 nm. The
films were patterned to micron scale wires with broad beam Ar ion
milling (1 mAcm$^{-2}$, 500 V), and then processed with a Ga focused ion
beam to achieve vertical transport with a device area in the
range 0.25 $\mu$m$^{2}$ - 1.2 $\mu$m$^{2}$. The fabrication process is
described in detail elsewhere \cite{bell}.  For the
bilayer measurements Cu/FeMn/Nb films were grown.

All transport measurements were made in a liquid He dip probe. The
critical current ($I_C$) and normal state resistance ($R_N$) were measured with
room temperature electronics. For devices with a $I_C R_N >
1$ $\mu$V a current-voltage ($I-V$) characteristic was directly
measured. For samples with $I_C R_N < 1$ $\mu$V a differential
resistance measurement was made with a lock-in amplifier, and the
$I_C$ was found as a peak in the d$V$/d$I$ measurement. For the $T_C$
measurements a four-point configuration was used on unpatterned
$5\times 10$ mm films. The rate of cooling was $< 10^{-2}$ K/s near $T_C$.

\section{Results}
In this Section the results of $T_C$ measurements on AF/S bilayers are
presented, followed by the measurements of S/AF/S Josephson
junctions.  
\subsection{Measurements of $T_C$ in Bilayers}
\begin{figure}[h]
\includegraphics[width=8cm]{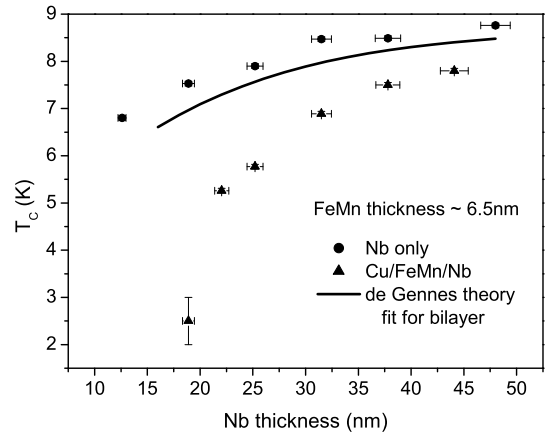}
\caption{\label{vary Nb}Variation of $T_C$ vs Nb thickness for a
constant FeMn thickness of 6.5 nm, (triangles), compared to plain Nb
films (circles). Fit to triangles is de Gennes theory (see Discussion).}
\end{figure}

\begin{figure}[h]
\includegraphics[width=8cm]{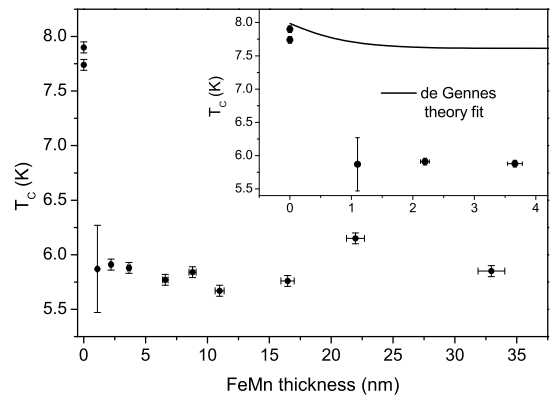}
\caption{\label{vary FeMn}Variation of $T_C$ vs FeMn thickness for a
constant Nb thickness of 25 nm. Inset: detail for thinner films, with
de Gennes theory fit, (see Discussion).}
\end{figure}

Following H\"ubener et al \cite{hubener} we have
measured the film $T_C$ independently varying the S and AF
thicknesses, as shown in Figs \ref{vary Nb} and \ref{vary FeMn}
respectively.  The $R(T)$ curves, showed a transition width of the
order of 0.1 K, and the $T_C$ is defined as the midpoint of the
transition. An absolute error of 0.05 K was found by measuring the
$T_C$ of a thick Nb film, and repeated $T_C$ measurements show a relative error of $\sim 0.05$ K. For
thicknesses of FeMn $< 1$ nm a broadened transition ($\sim 0.2$
K), is observed - presumably associated with large percentage
variation of the film thickness over the substrate, and no
consistent data was obtained in this thickness regime.

For a constant $d_{\mathrm{FeMn}} = 6.5$ nm, with varying Nb
thickness, a suppression of $T_C$ was observed relative to the plain Nb
film . In Fig. \ref{vary Nb} the final point with $T_C <
4.2$ K was measured in a closed cycle He-3 cryostat, with a
different calibration and thermal environment, hence the relative
error in $T_C$ is larger.
For a constant Nb thickness of 25 nm the $T_C$ of the film
drops dramatically as soon as the thinnest layer of FeMn is grown
underneath, (the two points at $d_{AF} = 0$ in Fig. \ref{vary
FeMn} are for Nb only and a Nb/Cu bilayer respectively). This
implies a short coherence length $\xi_{AF}$ in the FeMn- which we will show correlates
with the results from the Josephson junctions.
As a comparison, measurements on plain Nb films of decreasing
thickness were also made, since in this thickness regime, the Nb
only film $T_C$ is also decreasing, (circles in Fig. \ref{vary
Nb}). The reduction can be attributed to a combination of grain size and resistivity
effects, as well as the proximity effect \cite{minhaj}. The correlation between $T_C$
and resistance ratio R(295 K)/R(10 K), (RRR) shown in Fig. \ref{Nb
only} is associated with the grain size effect, and is consistent with previous studies \cite{andreone}.
\begin{figure}[h]
\includegraphics[width=8cm]{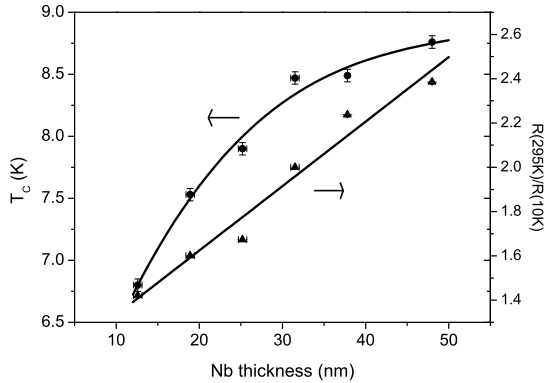}
\caption{\label{Nb only}$T_C$ and RRR variation vs thickness for
Nb only films. Lines are the best fit cubic and linear curves for
$T_C$ and RRR respectively.}
\end{figure}
\subsection{Josephson junctions}
\begin{figure}[h]
\includegraphics[width=8cm]{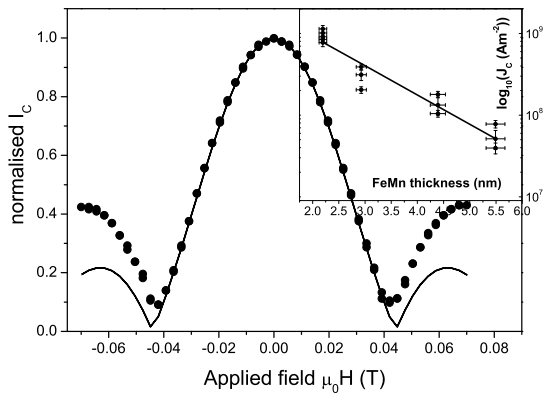}
\caption{\label{ICB}Critical current modulation with an
applied magnetic field, normalised to the zero field $I_C$. Line is a best-fit Fraunhofer pattern.
Inset: Critical current density vs FeMn thickness for junctions at 4.2
K. Line is a best fit exponential $\exp (-2d_{AF}/\xi_{AF})$ with $\xi_{AF}=2.4$ nm}
\end{figure}
The Josephson junctions showed resistively shunted junction (RSJ),
$I-V$ characteristics, with $I_C$ in the range 10 $\mu$A to 1.2 mA, 
and $R_N \le 2$ m$\Omega$. The re-entrant $I_C(H)$ in Fig.
\ref{ICB} shows the presence of a Josephson current through the
FeMn, although we do not obtain an ideal Fraunhofer pattern. In this
case the modulation is normalised to zero field critical current $I_C = 500$ $\mu$A. 
The junction dimension perpendicular to the direction of the
applied field was $\sim 600$ nm and the total barrier thickness, (Cu
and FeMn) was 7 nm. Correcting for the finite thickness of the Nb
electrodes \cite{weihnacht}, (each is 150 nm thick), we obtain from
the Fraunhofer fit a magnetic penetration depth of 40 nm.
A voltage criterion is used to extract the $I_C$, hence the non-zero
$I_C$ is an artifact of this process: the $I_C$ is suppressed to
zero to within the 1 $\mu$V noise level of the measurement. We can
estimate the Josephson penetration depth using
$\lambda_{J}=(\hbar/2e\mu_{0}J_{C}d)^{1/2}$. For the thinnest case
with $d_{AF} \sim 3$ nm  we have $J_C \sim 1 \times 10^{9}$
Am$^{-2}$ we find $\lambda_J \sim 2$ $\mu$m, whereas the largest
junction dimension is 1.2 $\mu$m, so we are close to long
junction behaviour only for the largest junction with the thinnest barriers.

The inset of Fig. \ref{ICB} shows the variation of critical current density,
$J_C$, with $d_{\mathrm{FeMn}}$. Assuming a dirty S/N/S junction $J_C \propto \exp (-k_{AF} d_{AF})$ \cite{degenRMP}. Using
$k_{AF} = 2 / \xi_{AF}$ (see Discussion) and fitting this
to the inset of figure \ref{ICB}, we find the characteristic decay length
$\xi_{AF} = 2.4$ nm. The errors in $J_C (d)$ consist of
measurement error of the sub-micron junction area, as well as
scatter due to variation of $d_{AF}$ over the area of the chip, (this
is also the case for the scatter in the $I_C R_N$ data of junctions
with nominally the same thickness of FeMn). There may also be
additional variation due to domain structures in the AF, and spin compensation at
the interfaces which are spatially inhomogeneous. All devices for a given
film thickness are patterned on the same 10 mm $\times$ 5 mm chip,
hence interface transparency and contamination should be comparable
for a given $d_{AF}$.

\section{Discussion}
$\gamma$-Fe$_{50}$Mn$_{50}$, (FeMn) is an example of a f.c.c.
disordered AF, and is one of the most studied materials used to
exchange bias ferromagnetic films \cite{nogues}. Although its
magnetic / exchange bias properties have been intensively studied, the
crystal structure is complex and difficult to probe, hence the
precise nature of the magnetic structure of this material is still 
under debate \cite{stocks2002,nakamurafemn}. The
electronic properties are less well studied, and the authors are
not aware of any experimental studies of properties such as the
Fermi velocity, electronic heat capacity, or mean free path of
FeMn thin films. In this Section we use the measurements above to
calculate some of the properties of the FeMn, and use them to
fit the bilayer $T_C$ measurements.

If we consider FeMn as a band antiferromagnet the coherence length
in the AF is  given by (in the dirty limit):
\begin{equation} \label{XiAF} \xi_{AF}= \left[ \frac{2\hbar D}{H_{Ex}}\right] ^{\frac{1}{2}} \end{equation}
where $D =\frac{1}{3}v_{Fermi}\ell$ is the diffusion constant, and
$H_{Ex} \sim k_B T_N$ the exchange coupling between the AF spins,
similar to the exchange field in the ferromagnetic case.
The additional factor of 2 compared to the F case ($\xi_F = (4 \hbar
D/H_{Ex})^{1/2}$) arises since the AF wave vector has the form $k_{AF} =
2/\xi_{AF}$ rather than the complex form
$k_F = 2(1 + i)/\xi_F$ which gives rise to the oscillating order
parameter in the F case \cite{Krivoruchko}. Hence, $\xi_{AF}$ scales with the bulk $T_N$, which in
the case studied here is in the range $450 - 490$ K
\cite{nogues}. In the case of FeMn, however, because it
is a highly disordered alloy system, we expect a very short mean
free path $\ell$, of the order of 1 nm or less \cite{rossiter}.
Hence, a reasonable value of $\xi_{AF}$ using equation (\ref{XiAF})
with a Fermi velocity $v_{fermi} = 2\times 10^6$ ms$^{-1}$ (both Fe
and Mn have $v_{Fermi} \sim
2\times 10^{6}$ ms$^{-1}$) is of the order of  $4 - 5$ nm.
This is in agreement with the value from the trilayer junctions,
and much shorter than the corresponding coherence length in a
normal metal. Further information can be gained from this
measurement of the coherence length. From equation (\ref{XiAF})
using $T_N$ = 450 K, and $\xi_{AF} = 2.4$ nm, the diffusion
constant $D$ is found to be $1.7 \times10^{-4}$ m$^2$s$^{-1}$.

\begin{figure}[h]
\includegraphics[width=8cm]{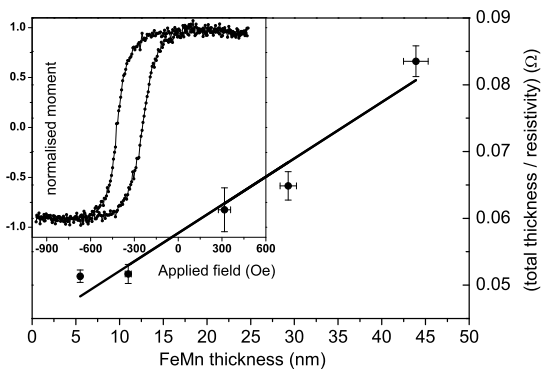}
\caption{\label{bias}Linear fit to find $\sigma_{FeMn}$ using
different thicknesses of FeMn in a Cu/FeMn/Nb trilayer, at 295
K. Inset: Hysteresis loop of a Nb/Cu/FeMn/Co/Nb trilayer after
annealing at $0.2$ T from 200 $^{\circ}$C for 30 minutes.}
\end{figure}

Given the value of the diffusion constant $D_{AF}$ it is possible to
calculate the density of states at the Fermi level using the Einstein relation $\sigma = 2e^2 \mathcal{N}
(\epsilon_F) D$, \cite{ashandmermin}. For this the value of $\sigma_{\mathrm{FeMn}}$ is required. This
was found using a series of Cu(5 nm)/FeMn/Nb(6 nm) films grown for differing FeMn thicknesses
(the Nb serves as a cap to prevent oxidation of the FeMn).
Assuming a simple parallel resistor model, plotting the ratio of total
thickness and resistivity against FeMn thickness should give a
straight line with gradient equal to the conductivity of the FeMn.
$\sigma_{\mathrm{FeMn}}$ is found to be $8.4 \times 10^5
(\Omega$m$)^{-1}$, from the linear fit (Fig. \ref{bias}). 
Hence using $\sigma = 2e^2 \mathcal{N}
(\epsilon_F) D$, with the values of $D$ and
$\sigma_{\mathrm{FeMn}}$ above, the density of states at the Fermi
level, $\mathcal{N}(\epsilon_F)$, in FeMn is found to be $2.7
\times 10^{25}$ states J$^{-1}$m$^{-3}$. For parallel resistors, the intercept of Fig.
\ref{bias} is given by $\sigma_{\mathrm{Nb}}d_{\mathrm{Nb}} +
\sigma_{\mathrm{Cu}}d_{\mathrm{Cu}}$. Using the value of
$\sigma_{\mathrm{Nb}}$ (see below) this intercept is predicted to be
of the order of 0.16 $\Omega$. The smaller observed value is associated
with interface resistance, which would be significant for
$d_{\mathrm{Nb}} = 6$ nm, which reduces the effective value of
$\sigma_{\mathrm{Nb}}$. This is also consistent with a higher value of
$\sigma_{\mathrm{FeMn}}$ 
obtained from the junction $R_N$
values: which would also contain a component due to the interface resistance.
We can now fit the bilayer $T_C$ measurements. We do not follow reference
\cite{hubener} which analysed the data in terms of the Werthamer
theory \cite{werthamer1963}. This considers the
case that the metals are identical in the normal state - i.e. the
Fermi velocities, residual resistivity and Debye temperatures are
the same, and uses a single effective coherence length. However in
this case the Nb and FeMn films are significantly different, and
we should consider the coherence lengths of the superconductor and
normal metal separately. Hence we use de Gennes theory in the one
frequency approximation \cite{degenRMP}. For the case that the normal metal film is
not superconducting at any temperature - i.e. has $T_C = 0$ K  we have:
\begin{eqnarray}\label{degennesTc} \frac{1}{\sqrt{2} \xi_S}\sqrt{\left(
\frac{T_{CS}}{T_C} -1 \right)} \tan \left[ \frac{d_S}{\sqrt{2} \xi_{S}}
\sqrt{\left( \frac{T_{CS}}{T_C} -1 \right)} \right] = \nonumber \\ 
 \frac{2}{\xi_{AF}}\frac{D_{AF} \mathcal{N}_{AF} (\epsilon_F)}{D_{S} \mathcal{N}_S (\epsilon_F)}
  \tanh \left[ \frac{2d_{AF}}{\xi_{AF}} \right],
\end{eqnarray}
where the largest root of this equation gives $T_C$, the transition
temperature of the bilayer. Here $\mathcal{N}_{S,AF}(\epsilon_F)$ are the density of
states at the Fermi level of the S and AF respectively, and $T_{CS}$
the transition temperature of the plain
S layer. All of the required parameters for equation (\ref{degennesTc})
are known, or can be obtained by experiment, as we now show.

From the trilayer junctions results above we use $\xi_{AF} = 2.4$
nm. The thicknesses $d_S$ and $d_{AF}$ are known for a given film.
To find $\xi_S$ in the dirty limit we use $\xi_S = 0.85(\xi_0
\ell)^{\frac{1}{2}}$, with $\xi_0 \approx \hbar v_{Fermi} / k_B T_{0}$
and substitute using the free electron form $D =
\frac{1}{3} v_{Fermi} \ell = \sigma k_B^2 \pi^2 /3 e^2 \gamma $, \cite{hubener, pippard}.
We obtain
\begin{equation}\label{xiequ}\xi_S = \left[ \frac{\pi \hbar k_B
\sigma}{6e^2 \gamma T_0}\right]^{\frac{1}{2}}.
\end{equation}
Here the electronic specific heat capacity $\gamma = 720$
Jm$^{-3}$K$^{-2}$, \cite{Poole} and $T_0 = 9.25$ K - the bulk $T_C$ of Nb. From a van der Pauw
measurement at 295 K we find a value of $\sigma_{\mathrm{Nb}} =
2.7\times 10^7$ $(\Omega$m$)^{-1}$ for our films. We use a linear fit
to follow the variation of RRR value with $d_{\mathrm{Nb}}$,
(Fig. \ref{Nb only}). Hence for a given thickness of Nb we
calculate $\sigma_{\mathrm{Nb}}$(10 K), using the RRR linear fit,
and then find $\xi_S$ using equation (\ref{xiequ}). $\xi_S$ is found to
be of the order of 6 nm. 

The plain Nb transition temperature $T_{CS}$ is similarly followed
using an empirical cubic fit to the $T_C$, as in
Fig. \ref{Nb only}. Finally from $\sigma = 2e^2 \mathcal{N}
(\epsilon_F) D$ the ratio $ D_{AF}\mathcal{N}_{AF} (\epsilon_F) /
D_{S}\mathcal{N}_S  (\epsilon_F)$ is identical to
$\sigma_{\mathrm{FeMn}} / \sigma_{\mathrm{Nb}}
 = 0.031$, using the above values. Hence the only
remaining parameter in equation (\ref{degennesTc}) is $T_C$, for
which we solve numerically.

For varying Nb thickness, this fit with no adjustable parameters
is shown as a solid line in Fig. \ref{vary Nb}, which is clearly not a
good qualitative fit to the data. For the case of varying FeMn thickness, as can be
seen from Fig. \ref{vary FeMn}, the suppression of $T_C$ is
saturated for thicknesses $>$1 nm: which would be expected for a
coherence length of that order. The inset of Fig. \ref{vary FeMn} 
shows that the theoretical fit, which saturates for $d_{AF}
\ge 2.5$ nm, although the saturation value of $T_C$ is much
higher than found experimentally. With $d_{\mathrm{FeMn}} > 10$ nm
there does appear to be some additional variation of $T_C$, which is
not expected from equation (\ref{degennesTc}). However this most
likely due to a different phase of FeMn being produced in films
thicker than 20 nm \cite{tsang}. More investigation of this behaviour
is required. 

A Nb(150 nm)/Cu(5 nm)/FeMn/Co/Nb(150 nm) structure was also grown
as a reference to check the AF nature of the FeMn, with
$d_{\mathrm{FeMn}} \sim 4.5$ nm, and $d_{\mathrm{Co}} \sim 2$ nm.
$M(H)$ hysteresis loop measurements were made with a vibrating sample
magnetometer, and showed that there was some exchange bias associated with the applied
field during deposition. The relatively weak nature of this
($H_{bias} \sim 150$ Oe) implies that there are many misaligned
domains in the Co being pinned by the AF. The film was annealed
in a field of $0.2$ T at $200$ $^{\circ}$C for 30 minutes, and
field cooled. After annealing the exchange bias was measured as
$H_{bias} \sim 335$ Oe, (Inset of Fig. \ref{bias}). This shows that the
FeMn is an AF in this thickness regime, as expected.
Although in this regime the magnitude of the exchange bias is changing
with $d_{\mathrm{FeMn}}$ \cite{jungblut}, this does not imply
variation of the exchange coupling energy between the spins at 4.2 K, hence we can assume a
constant value of $\xi_{AF}$.

Many models of exchange bias in magnetic films use compensation of
spins at the interface as a crucial parameter. Indeed the true
structure of the present device might be S/N/F/AF/F/S: where the F
layers are uncompensated AF spins. A full theoretical description
of these bilayers and junctions may enable additional information
concerning the nature of thin films of $\gamma$-FeMn to be gained, as
well as provide a qualitative fit to the data in the present work.

\section{Conclusions}
We have been able to obtain the diffusion constant $D$, and
density of states of $\gamma$-FeMn. We have also fabricated AF
Josephson junctions with a coherence length $\xi_{AF} \sim 2.4$ nm.
This value of $\xi_{AF}$ was used to model AF/S bilayer $T_C$
measurements, and with no free parameters, but gave unsatisfactory
results. It is clear that a more relevant theory is required to fit
the data. Although this work implies a similar strong suppression
of superconductivity in AF FeMn as in the ferromagnetic case, the
thicknesses used in the present work are similar to those used to
fabricate spin-valves. Hence the fabrication of a Josephson
spin-valve using FeMn as a pinning layer, measured below 4.2 K may
possess a $J_C$ which is not too small to be beyond experimental
reach, although a AF with lower $T_N$, and weaker FM layers would
significantly increase the $J_C$ of the junctions.
\begin{acknowledgments}
We would like to thank Manna Ali (University of Leeds) for annealing the exchange biased film.
This work is supported by the Engineering and Physical Sciences
Research Council, UK.
\end{acknowledgments}

\end{document}